\documentclass[floatfix,amsmath,amssymb,aps,prl,twocolumn]{revtex4}
\usepackage{graphicx}

\newcommand{\veps}{\varepsilon} 

\newcommand{\me}{m_{\mathrm{e}}}

\newcommand{\lC}{\lambda_{\mathrm{C}}}

\newcommand{\dd}{{\mathrm{d}}}

\newcommand{\Beta}{{\mathrm{B}}}

\newcommand{\SPvec}[1]{{\textbf{\textsl{#1}}}}

\newcommand{\haelfte}{{\textstyle{\frac{1}{2}}}}
\newcommand{\viertel}{{\textstyle{\frac{1}{4}}}}

\newcommand{\hfrac}[2]{{{#1}/{#2}}}

\newcommand{\phiBI}{{\phi_{\beta}}}

\begin{document}
%
%
	\title{Nonperturbative calculation of Born-Infeld effects on 
		the Schr\"odinger spectrum of the hydrogen atom} 
%
%
%
\author{Holly Carley}
\email{carley@math.rutgers.edu}
\affiliation{Rutgers University, 
	Department of Mathematics, 
	110 Frelinghuysen Rd., 
	Piscataway, NJ 08854}
\author{Michael K.-H. Kiessling}
\email{miki@math.rutgers.edu}
\affiliation{Rutgers University, 
	Department of Mathematics, 
	110 Frelinghuysen Rd., 
	Piscataway, NJ 08854}

%
%

\begin{abstract}
We present the first nonperturbative calculations of 
the nonrelativistic hydrogen spectrum as predicted by 
first-quantized nonlinear Maxwell-Born-Infeld electrodynamics with
point charges.
 Judged against empirical data our results 
significantly restrict the range of viable values of the 
new electromagnetic constant $\beta$ introduced by Born.
 We assess Born's own proposal for the value of $\beta$.
\end{abstract}
\maketitle

%
%

In the twenty years since its rediscovery in the 26-dimensional bosonic 
string theory study by Fradkin and Tseytlin \cite{efat}, the nonlinear 
electromagnetic field theory proposed by Born and Infeld \cite{mbliA}
has been experiencing an astonishing renaissance.
Recent surveys are \cite{ggB,ggA}. 
 Most investigations since  \cite{efat} have been 
conducted from the perspective of the high energy community and involve 
higher-dimensional versions of the Born-Infeld theory (as in 
\cite{efat,yy}) 
and/or non-commutative analogs of it (as in \cite{dgrk}).  
  Inevitably this has rekindled the interest in the original 
four-dimensional theory, the subject of this letter.

 We recall that Born's agenda \cite{mbA} was to rid (early)
QED from its ultraviolet divergencies by quantizing 
self-regularizing nonlinear classical field equations.
 It was noted already in \cite{mbA} that the nonlinear 
Maxwell-Born-Infeld field equations 
\cite{remarkA}
{In \cite{mbA} Born proposed a simpler 
	nonlinearity than in \cite{mbliA}; yet in the electrostatic 
	limit both field theories coincide.}
do not lead to the infinite self-energy problems of a 
point charge which occur with the linear Maxwell-Lorentz 
field equations, but the nonlinearity made it difficult 
to proceed. 
 With the spectacular quantitative successes 
of renormalized QED since the late 1940s, Born's original 
motivation became obsolete; or so it would seem. 
 However, as emphasized by Weinberg \cite{swA}, more than half a century 
later standard QED is still in need of
extrinsic mathematical regularizers to overcome the infinite self-energy
problems of a point charge that have been inherited, in a sense, from
the classical Maxwell-Lorentz electrodynamics.
 In view of this, Born's suggestion \cite{mbA} to pursue some 
intrinsically self-regularizing nonlinear electromagnetic field 
theory reads as contemporary as it did in the 1930s;
the rediscovery of Born-Infeld type Lagrangians in string theory, 
which could hardly have been foreseen by its founders, 
makes Born's suggestion all the more prophetic.

 The avoidance of infinite self-energies, as well as some other conceptual 
items \cite{ibb}, are greatly to the theory's credit but surprisingly 
little is known about the empirical validity of the Born-Infeld theory. 
 While the theory does not seem to have problems at
the classical level \cite{es,mkA} it remains to be seen
whether it will live up to its expectations at the quantum level.

 In this vein, a very natural question to ask is the following:
\emph{What (detectable) effects 
does a hypothetical Born-Infeld nonlinearity of the electromagnetic 
fields have on the atomic spectra?}
 This question should have been answered long ago.  
 It was not, presumably because the nonlinearity of the field
equations causes
``difficulties [...] with the passage to the quantum theory, which 
appear to be insoluble with present methods of quantization''
\cite{pamd} (p.32), and by 1969
``[t]he adaption [...] to the principles of quantum theory and the
introduction of the spin ha[d] [...] met with no success''
\cite{mbB} (p.375).
 As long as this situation prevails, one has to settle for 
quantum \emph{mechanical} computations of spectral data in which 
Born-Infeld effects can be incorporated through the classical fields.

 Unfortunately, because the complicated nonlinearity of the 
field equations has stood in the way of finding relevant 
solutions with two or more point charges, all previous attempts 
to compute such quantum mechanical spectra 
\cite{ghlm}, \cite{jrlfwg}, \cite{gsjrwg}
have been foiled.
 In 
    \cite{ghlm}
the electron is treated as a test particle in the known
	(see \cite{mbA})
Maxwell-Born-Infeld field of a point nucleus to compute hydrogen-like
Schr\"odinger spectra to first order in perturbation theory; however, 
as we will see in this article, test particle theory is misleading for 
Born-Infeld equations.
 In 
  \cite{jrlfwg} and \cite{gsjrwg}, 
which have become standard references
(see the introductions in \cite{gmr} and \cite{jdj}),
Dirac spectra are computed without recourse to test particle theory 
(albeit with other approximations which are not of concern here),
defining the interaction energy as difference of the electrostatic field 
energy integrals for the bound versus the free configurations.  
 However, the authors of 
  \cite{jrlfwg} and \cite{gsjrwg}, who
use Coulomb's solution $\SPvec{D}_{\mathrm{C}}$ of the displacement
field equation $\nabla\cdot\SPvec{D} = 4\pi\rho$ with a charge density
$\rho$ comprising a single spectral electron and a spherically symmetric 
nucleus of charge $z$ and a Thomas-Fermi cloud for the remaining $z-1$ 
electrons, fail to realize that the 
nonlinear Born-Infeld law for the electromagnetic vacuum maps this 
Coulomb field $\SPvec{D}_{\mathrm{C}}$ into an electric field 
$\SPvec{E}_{\mathrm{FGRS}} = {\cal F}_{\mathrm{BI}}(\SPvec{D}_{\mathrm{C}})$ 
which is not identically curl-free 
\cite{remarkB};
more precisely,
$\nabla\times {\cal F}_{\mathrm{BI}}(\SPvec{D}_{\mathrm{C}}) \neq \SPvec{0}$ 
almost everywhere, invalidating the spectral results of \cite{jrlfwg} 
and \cite{gsjrwg}.

 Recently, a consistent first quantization of the nonlinear 
Maxwell-Born-Infeld field equations with point charges 
was achieved using the electromagnetic potentials \cite{mkB}.
 Moreover, an explicit integral formula for the electron's electrostatic 
potential in certain proton-electron configurations (treated as point charges) 
was derived; this integral formula is readily extended to nuclear charges 
$z>1$ (see below).
 Thus the stage has been set for a systematic investigation of 
the simplest atomic and ionic spectra, the hydrogen-like ones.

 In order to keep technical matters as simple as possible, here we
only address the non-relativistic Schr\"odinger equation of a spinless 
electron bound to an infinitely massive point nucleus. 
 We plan to deal with the fine details contributed by relativity, 
spin, and the finite mass and size of the nucleus elsewhere.
 Furthermore, detailed evaluations of the interactions and the
eigenvalues are carried out only for the hydrogen atom ($z=1$);
the details of hydrogen-like interactions and ionic spectra for 
nuclear charges $z>1$ are beyond the scope of this letter.

 In units of $\hbar$ for both action and 
magnitude of angular momentum, elementary charge $e$ for charge, 
electron rest mass $\me$ for mass,
speed of light $c$ for velocity,
and Compton wave length of the electron $\lC = {\hbar}/{\me}c$ 
for both length and time, a hydrogen-like spectrum is determined by the 
following dimensionless stationary Schr\"odinger equation
on the electron's configuration space \cite{mkB},
\begin{equation}
- \haelfte\nabla^2_e \psi(\SPvec{s}_e)
- \alpha {\phiBI}(\SPvec{s}_e)\psi(\SPvec{s}_e)
=
E \psi(\SPvec{s}_e)
\,,
\label{eq:SCHROEDINGEReqHYDROGEN}
\end{equation}
where $\SPvec{s}_e$ is the electron's \emph{generic} configuration
space coordinate and the subscript $_e$ on $\nabla_e^2$ indicates
differentiation with respect~to~$\SPvec{s}_e$.
 The fine structure constant
$ 
\alpha 
\equiv 
\hfrac{e^2}{{\hbar}c}
\approx \hfrac{1}{137.036}
\, 
$
is the dimensionless electromagnetic coupling constant 
for the dimensionless \emph{total electrostatic potential}
$\phiBI$ defined below.
 The positive parameter $\beta$ is 
\emph{Born's electromagnetic vacuum constant}
(``aether constant'' for short)
\cite{remarkC},
which enters through the Born-Infeld aether law, relating $\SPvec{E}$
(and $\SPvec{H}$) with $\SPvec{D}$ (and $\SPvec{B}$).
 Born  \cite{mbA} argued that $\beta={\beta}_{\mathrm{B}}$ with
\begin{equation}
{\beta}_{\mathrm{B}}
\approx 
1.2361{\alpha}
\,.
\label{eq:BORNconstant}
\end{equation}
  Our spectral results allow us to
assess the viability of (\ref{eq:BORNconstant})
and Born's reasoning for it.

 The total electrostatic potential $\phiBI(\SPvec{s})$
at the actual space point ${\SPvec{s}}$ is determined by
the electrostatic Maxwell-Born-Infeld equation 
$\nabla\cdot {\cal F}^{-1}_{\mathrm{BI}}(-\nabla\phiBI)  = 4\pi\rho$ 
with $\rho$ consisting of one positive and one negative point charge 
with values $z$ and $-1$ at generic positions $\SPvec{s}_n$ and $\SPvec{s}_e$,
respectively; explicitly, 
\begin{equation}
-\nabla \cdot
   \frac{\nabla{\phiBI}(\SPvec{s})}
	{\sqrt{1-\beta^4|\nabla{\phiBI}(\SPvec{s})|^2 }}
=
4\pi \left(z\delta_{{\scriptstyle\SPvec{s}}_n}(\SPvec{s})
-\delta_{{\scriptstyle\SPvec{s}}_e}\!(\SPvec{s})\right),
\label{eq:PHIstatic}
\end{equation}
with the asymptotic condition that ${\phiBI}({\SPvec{s}})\to 0$
for $|{\SPvec{s}}|\to\infty$.
 The solution of (\ref{eq:PHIstatic}) depends on ${\SPvec{s}}$
as variable and on $\SPvec{s}_n$ and $\SPvec{s}_e$ as parameters;
we sometimes emphasize this by writing
$\phiBI(\SPvec{s}{}|\SPvec{s}_n,\SPvec{s}_e)$.
 While no explicit formula for $\phiBI(\SPvec{s}{}|\SPvec{s}_n,\SPvec{s}_e)$
is known, (\ref{eq:SCHROEDINGEReqHYDROGEN}) reveals
that we need to know only $\phiBI(\SPvec{s}_e|\SPvec{s}_n,\SPvec{s}_e)$.
 Fortunately, although
$\nabla\times {\cal F}_{\mathrm{BI}}(\SPvec{D}_{\mathrm{C}}(\SPvec{s})) 
\neq \SPvec{0}$ for almost every $\SPvec{s}$ in space, we do have
$\nabla\times {\cal F}_{\mathrm{BI}}(\SPvec{D}_{\mathrm{C}}(\SPvec{s})) 
= \SPvec{0}$ 
\emph{for all $\SPvec{s}$ on the straight line through the point charges}
(this result generalizes to the vanishing of 
$\nabla\times{\cal F}_{\mathrm{BI}}(\SPvec{D}_{\mathrm{C}}(\SPvec{s}))$ 
on the straight line through the respective centers of any two spherically 
symmetric charge distributions). 
 Hence, an electrostatic potential function $\phi(\SPvec{s})$ 
solving (\ref{eq:PHIstatic}) 
for space points $\SPvec{s}$ on that line 
\emph{can} be computed through 
the line integral 
$\phi (\SPvec{s}) = 
\int_{\SPvec{s}}^\infty 
{\cal F}_{\mathrm{BI}}(\SPvec{D}_{\mathrm{C}}(\SPvec{s}^\prime)) 
\cdot d\SPvec{s}^\prime$. 
 Assuming $\SPvec{D} = \SPvec{D}_{\mathrm{C}}$ 
in leading order in $\beta$, on this line we can approximately set
$\phi (\SPvec{s}_e) = \phiBI (\SPvec{s}_e)$. 
 For $z>1$ the integral is  formidable, but
when the nucleus is a proton ($z=1$ and $\SPvec{s}_n = \SPvec{s}_p$),
it can be recast into the more managable form  \cite{mkB}
\begin{equation}
\phiBI(\SPvec{s}_e|\SPvec{s}_p,\SPvec{s}_e) 
=
\frac{1}{ \beta}
\int_{2\sqrt{2}\beta/r}^{\infty}
\frac{f^\prime(y)}{\sqrt{1+ x^4}}
      \dd{x}
\,,
\label{eq:AnullATsONElargeINT}
\end{equation}
where $r = |\SPvec{s}_p-\SPvec{s}_e|$, 
$xy = \beta/r$, and $f^\prime$ is the derivative of 
\begin{equation}
f(y) = \sqrt{\viertel + y^2 - y\sqrt{1 + y^2}}
\,.
\label{eq:fDEFINITION}
\end{equation}
  For the remainder of this letter, $z=1$.

 A look at the integral (\ref{eq:AnullATsONElargeINT})
makes it plain that $\beta\phiBI(\SPvec{s}_e|\SPvec{s}_p,\SPvec{s}_e)$ 
depends on $\SPvec{s}_p$, $\SPvec{s}_e$ and $\beta$ only through the 
combination $|\SPvec{s}_p-\SPvec{s}_e|/\beta$;  
hence, 
$\beta\phiBI(\SPvec{s}_e|\SPvec{s}_p,\SPvec{s}_e) =: W(r/\beta)$
is a function of $r/\beta$.
 And while $W$ does not seem to be expressible in terms of known 
functions,  (\ref{eq:AnullATsONElargeINT})  
lends itself readily to an analysis when the electron is far from,
respectively near the proton. 
 Note that ``far'' and ``near'' are relative to $\beta$.

 If the electron is far from the nucleus, i.e., if
$r \geq 2\sqrt{2}  \beta$, then $W(r/\beta)$ can be 
expanded in an asymptotic series in powers of $\beta/r$
(asymptotically exact to four orders as $r \to \infty$), 
thus
\begin{equation}
W(r/\beta)
=
\sum_{k=0}^3 b_k\left(\hfrac{\beta}{r}\right)^{k}
+ o\left((\hfrac{\beta}{r})^3\right) 
\,,
\label{eq:AatSoneASYMPTOTICS}
\end{equation}
with
$b_0 =  -\frac{1}{4}B\left(\frac{1}{4},\frac{1}{4}\right)$,
$b_1 = 1$, 
$b_2 = \frac{3}{4}B\left(\frac{3}{4},\frac{3}{4}\right)$,
and
$b_3 = 2$, 
where B$(\,.\,,\,.\,)$ is Euler's Beta function.
 Formula (\ref{eq:AatSoneASYMPTOTICS}) reveals three important 
results.
 First, when $r\to\infty$ the electric potential 
at the location of the electron, $\phiBI(\SPvec{s}_e)$,
converges  to the \emph{finite} electron self-potential in Born-Infeld 
theory, defined by setting $\SPvec{s}=\SPvec{s}_e$ in Born's solution 
\begin{equation}
\phi_{\mathrm{B}}^{(-)}({\SPvec{s}}{}|{\SPvec{s}_e}) 
=  
-\frac{1}{\beta}
\int_{|{\scriptstyle{\SPvec{s}}-\SPvec{s}_e}{}|/\beta}^\infty 
\frac{\dd{x}}{\sqrt{1+ x^4}} 
\,
\label{eq:BornsElectricPot}
\end{equation}
for the electrostatic potential at $\SPvec{s}$ 
generated by a single (negative) unit point charge 
at~$\SPvec{s}_e$~\cite{mbA},
(NB: (\ref{eq:BornsElectricPot}) solves (\ref{eq:PHIstatic}) when 
$z = 0$).
We recall \cite{mbA}: 
\emph{there is no short distance Coulomb singularity of the single
	particle potential in Born-Infeld theory}.
 Second, to leading order for large separation of electron and proton, 
the potential $\phiBI(\SPvec{s}_e)$ \emph{varies} with $r$ reciprocally,
i.e., we recover \emph{Coulomb's  law for the pair potential} from 
the nonlinear field equation (\ref{eq:PHIstatic}). 
 Third, there are higher order corrections to Coulomb's law.
 Indeed, when the electron is near the nucleus, deviations from Coulomb's law 
become significant.
 More precisely, for $r < 2\sqrt{2}  \beta$, 
the function $W(r/\beta)$ can be expanded into a Taylor series in 
powers of $r/\beta$, 
\begin{equation}
W(r/\beta)
= 
\sum_{k=0}^\infty a_k\left(\hfrac{r}{\beta}\right)^{4k+1}
\,
\label{eq:AatSoneSMALLsONE}
\end{equation}
with explicitly computable expansion coefficients $a_k$.
 The first four of them read as follows: $a_0 = -1/2$, 
$a_1 = 3/40 -3\pi/138$, 
$a_2 = -29/672+ 225\pi/16384$,
and 
$a_3 = 1667/54912 -20265\pi/2097152$.
 Note in particular that $W(0)=0$: \emph{there is no short distance
Coulomb singularity of the pair potential in Born-Infeld theory} \cite{mkB}.

 Our discussion of $\phiBI$ supplies all the information we need to 
solve the Schr\"odinger equation (\ref{eq:SCHROEDINGEReqHYDROGEN}).
 To facilitate the comparison with the familiar
Schr\"odinger equation for the Coulomb interaction, we 
write the eigenvalues as 
$E= 
\frac{\alpha}{\beta}\frac{1}{4}B\left(\frac{1}{4},\frac{1}{4}\right)+\veps$
and the total potential as
$\phiBI(\SPvec{s}_e) = 
- \frac{1}{\beta}\frac{1}{4}B\left(\frac{1}{4},\frac{1}{4}\right) +
\frac{Z(r/\beta)}{r}$, where $Z(r/\beta)$ is the 
\emph{effective Coulomb charge} of the proton ``seen'' from a distance $r$.
 The self-potential terms on left and right hand side of 
(\ref{eq:SCHROEDINGEReqHYDROGEN}) then cancel out, leaving us to solve
\begin{equation}
- \haelfte\nabla^2_e \psi(\SPvec{s}_e) 
- \alpha {\textstyle{\frac{Z(r/\beta)}{r}}}
\psi(\SPvec{s}_e) 
=
\veps \psi(\SPvec{s}_e) 
\,.
\label{eq:SCHROEDINGEReqHYDROGENnormed}
\end{equation}
\begin{figure}\label{SCHROEpots}
\includegraphics[height=4cm]{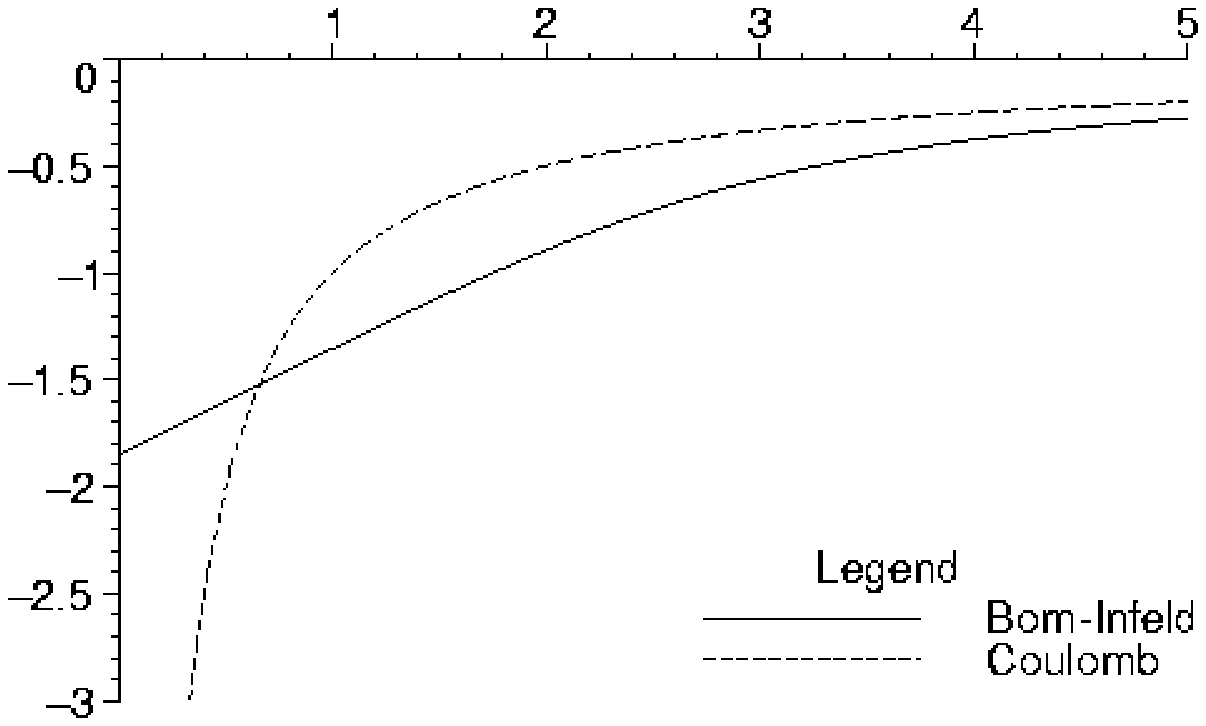}
\caption{ $-(\hfrac{\beta}{r})Z(\hfrac{r}{\beta})$ and 
$-\hfrac{\beta}{r}$ vs. $\hfrac{r}{\beta}$.
\hfill}
\end{figure}
  The Born-Infeld Schr\"odinger and the Coulomb Schr\"odinger potentials 
are compared in Fig.~1.

 Our first important spectral result states that the Coulomb limit 
$\beta\to{0}$ of eq. (\ref{eq:SCHROEDINGEReqHYDROGENnormed}) exists.
 In this case, $Z(r/\beta) \to 1$ for all $r>0$,
which follows from (\ref{eq:AnullATsONElargeINT}) \cite{mkB}.
 Hence, in the limit $\beta\to 0$
the spectrum of (\ref{eq:SCHROEDINGEReqHYDROGENnormed}) 
converges to the familiar Rydberg law, i.e., 
\begin{equation}
 \veps_{n,\ell,m}^{(0)}  = - \textstyle{\frac{1}{2n^2}}\alpha^2
,\qquad n = 1,2,...
\,,
\end{equation}
where $n=1,2,3,...$ and $\ell=0,1,...,n-1$ and $m=-\ell,...,0,...,\ell$
are the usual main, secondary, and magnetic quantum numbers.
  As is well known, $\ell$ and $m$ do not contribute to the energy 
eigenvalues $\veps^{(0)}_{n,\ell,m}$, so that $n^2$ of them coincide;
we recall that this high degeneracy is due to the $O(4)$ invariance
of (\ref{eq:SCHROEDINGEReqHYDROGENnormed}) when $\beta=0$.
 
 For all $0<\beta<\infty$, the $O(4)$ invariance is broken and 
the energy eigenvalues $\veps_{n,\ell,m}^{(\beta)}$ in general display 
only the $2\ell+1$-fold degeneracy corresponding to the manifest $O(3)$ 
invariance of (\ref{eq:SCHROEDINGEReqHYDROGENnormed}); i.e., 
$\veps_{n,\ell,m}^{(\beta)}$ does not depend on the quantum number $m$.
The $O(3)$ symmetry allows us to treat
(\ref{eq:SCHROEDINGEReqHYDROGENnormed}) by the usual
separation of variables.
 Shifting the origin of space to $\SPvec{s}_p$, the electron-proton
 distance $r$  becomes  the radial variable of 
standard spherical coordinates $r,\vartheta,\varphi$. 
 In these coordinates the eigen-wavefunctions take the form
$\psi_{n,\ell,m}^{(\beta)} (\SPvec{s}_e) 
= 
R_{n,\ell}^{(\beta)}(r)Y_\ell^m(\vartheta,\varphi)$, where
the $Y_\ell^m(\vartheta,\varphi)$ are spherical harmonics, 
and the $R_{n,\ell}^{(\beta)}(r)$ satisfy the Sturm-Liouville
problem
\begin{equation}
({r}^{2} R^\prime)^\prime
\!-\![ \ell(\ell\!+\!1) 
\!-\! 2\alpha rZ(r/\beta)  
\!-\!2\veps r^2 ]R =0 
\label{linRode}
\end{equation}
for $\int_0^\infty r^2R^2(r)dr<\infty$.
 We solved this radial problem by standard 
shooting technique, using MAPLE's Runge-Kutta-Fehlberg45 method.

 It is instructive to discuss first the dependence of the
ground state energy $\veps_0(\beta) \equiv \veps^{(\beta)}_{1,0,0}$
on $\beta$.
 In Fig.~2 we display our numerically computed values of 
$\veps_0(\beta)$ for a selection of $\beta$ values vs. $\beta$,
together with semi-explicit upper and lower bounds 
on $\veps_0(\beta)$, computed analytically 
except for numerical quadratures.

\begin{figure}\label{EnullOFbeta}
\includegraphics[height=5cm]{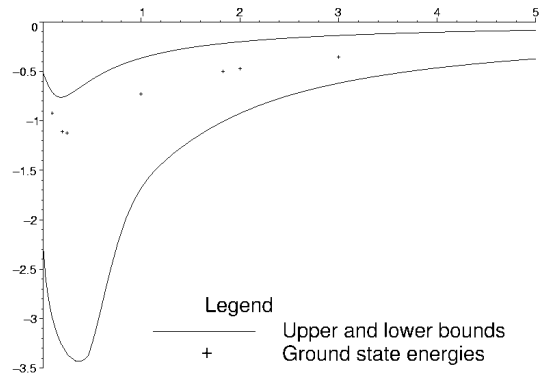}
\caption{ $\veps_0/\alpha^2$ vs. $\alpha\beta$.\hfill}
\end{figure}
 There are several remarkable features visible in Fig.~2.
 First of all, there are \emph{two} values of $\beta$ at which
the ground state energy $\veps_0(\beta)$ coincides with the familiar 
Coulomb value $-\alpha^2/2$.
 Thus there are two regimes where the Born-Infeld theory yields
a \emph{binding (or ionization) energy} ($-\veps_0(\beta)$) compatible 
with the empirical data:
a perturbative one near $\beta=0$ (the obvious one, as already noted) 
and a highly non-obvious and non-perturbative regime near
$\beta \approx 1.83297/\alpha$. 
 Too far away from these two values, the binding energy would be either
unrealistically large or small. 
 Between these two $\beta$ values, the binding energy
is enhanced compared to $\alpha^2/2$, 
reaching a maximum at about $\beta\approx 0.24774/\alpha$, while
to the right of $\beta \approx 1.83297/\alpha$, the binding energy 
is diminished, converging to zero as $\beta\to\infty$.
 We remark that our semi-explicit upper and lower bounds 
allow us to rigorously prove this nonmonotonic behavior of the binding 
energy.

 We emphasize that the nonmonotonic behavior of $\veps_0(\beta)$ is a
quite nontrivial result; in particular, there is no hint of it when the 
electron is treated (in ``first approximation'') as a \emph{test particle} 
which ``feels'' only the proton's electrostatic potential computed from the 
Born-Infeld equations with a single point source, neglecting the electron's 
own feedback
    \cite{ghlm}; 
i.e., $\phiBI(\SPvec{s}_e)\equiv \phiBI(\SPvec{s}_e|\SPvec{s}_p,\SPvec{s}_e)$ 
in (\ref{eq:SCHROEDINGEReqHYDROGENnormed}) is 
replaced by Born's \cite{mbA} solution for a positive point charge,
$\phi_{\mathrm{B}}^{(+)}({\SPvec{s}_e}{}|{\SPvec{s}_p})
=
-\phi_{\mathrm{B}}^{(-)}({\SPvec{s}_p}{}|{\SPvec{s}_e})$.
 For $r$ large,
$\phi_{\mathrm{B}}^{(+)}({\SPvec{s}_e}|{\SPvec{s}_p}) \sim \hfrac{1}{r}$, 
too.
 But since $\sqrt{1+x^4} > x^2$, it follows from 
(\ref{eq:BornsElectricPot}) that 
$\phi_{\mathrm{B}}^{(+)}({\SPvec{s}_e}|{\SPvec{s}_p}) < \hfrac{1}{r}$,
so \emph{test particle theory predicts 
a diminished binding energy for all $\beta$}.

 For a judicious selection of $\beta$ values we computed several 
higher eigenvalues.
 Of particular interest are Born's value (\ref{eq:BORNconstant}) and
the value $\beta \approx 1.83297/\alpha$ where the binding energy
coincides with the Coulomb value $\alpha^2/2$ at $\beta=0$.
 Listed in the table below are the energies of the ground, the first 
excited $s$, and the first $p$ states for these $\beta$ values,
as well as the corresponding empirical data \cite{data}.
 We display $\alpha\beta$ rather than $\beta$, and 
$-\veps/\alpha^2$ rather than $\veps$; also,
we suppress the magnetic quantum number $m$:

\begin{eqnarray*}
\nonumber
\begin{matrix}
\alpha\beta 
& -\veps_{1,0}^{(\beta)}/\alpha^2 
& -\veps_{2,0}^{(\beta)}/\alpha^2 
& -\veps_{2,1}^{(\beta)}/\alpha^2 \cr
 0.0000               & 0.50000 & 0.12500 & 0.12500 \cr
 6.6\!\times\!10^{-5} & 0.50016 & 0.12502 & 0.19101 \cr
 1.83297              & 0.50000 & 0.19766 & 0.36737 \cr
empirical	      & 0.49973 & 0.12493 & 0.12493 
\end{matrix}
\end{eqnarray*}

 From the table and Fig.~2 we are able to delineate the
physically viable range of $\beta$ values.
 By inspecting the excited energies, we can immediately rule out 
$\beta \approx 1.83297/\alpha$, and by continuity also its neighborhood.
 That leaves only the perturbative regime of sufficiently small $\beta$.
 But how small is ``sufficiently small''? In particular, 
is $\beta_{\mathrm{B}}
$  ``small enough''? 

 The second row in the table lists spectral data for 
$\beta=\beta_{\mathrm{B}}$.
 Note that $-\veps_0(\beta_{\mathrm{B}})/\alpha^2$ 
deviates from $-\veps_0(0)/\alpha^2=1/2$ (first row) by $1.6\times 10^{-4}$, 
and from the empirical data even by $4.3 \times 10^{-4}$. 
 Of course, $-\veps_0(0)/\alpha^2$ differs itself from the empirical 
data by $2.7\times 10^{-4}$, but as is well known, after correcting
$-\veps_0(0)/\alpha^2$ for the finite mass of the proton, the 
difference to the empirical data reduces to only $-3.14\times 10^{-6}$, 
and the agreement improves even more with relativistic corrections.
 It is therefore to be expected that even after correcting 
$-\veps_0(\beta_{\mathrm{B}})/\alpha^2$ for the finite
proton mass and relativistic effects, the difference to empirical
data will remain at about $10^{-4}$. 

 Even more dramatic is the splitting of $0.066\alpha^2$ between the
$2s$ and $2p$ energies computed with (\ref{eq:BORNconstant}), which 
is~a~factor $10^4$ bigger than the $2p_{3/2}-2p_{1/2}$ ``fine structure'' 
(which is not even visible at the level of precision in our 
table).
 Hence, 
even from the spectrum of the simplest atom we conclude that
{Born's value (\ref{eq:BORNconstant}) is not physically viable!}

 Pending verification of our results through a more refined treatment, 
viable values of $\beta$, as far as spectral results go, must be 
much smaller than $\alpha$.
We plan to study just how small $\beta$ must be using a relativistic 
theory with spin.
 But would the elimination of (\ref{eq:BORNconstant}) not be a bearer of
bad tidings for the Born-Infeld theory? 
 Not yet! 
 Born did \emph{not} use \emph{detectable energy differences}
to compute (\ref{eq:BORNconstant}) but equated the static field energy 
of a point charge to the electron's empirical rest energy, which yields
\begin{equation}
{\textstyle\frac{1}{4\pi}
\frac{\alpha}{\beta^4}}\!
\int\! (
{\sqrt{ 1 - \beta^4|{\nabla\phi_{\mathrm{B}}}|^2 }}^{\,-1}\! -\! 1)
d^3{\SPvec{s}}
=
1
\label{eq:HfuncFIELDS}
\end{equation}
in our units of ${\me}c^2 =1$.
 The integral equals ${\textstyle\frac{\alpha}{\beta}\frac{1}{6}}
   \Beta\left({\textstyle{\frac{1}{4},\frac{1}{4}}}\right)$,
giving (\ref{eq:BORNconstant}).
 Unless (\ref{eq:HfuncFIELDS})
can be tied to a \emph{dynamical} concept, such as scattering
of an electron, the elimination of (\ref{eq:BORNconstant}) 
by our spectral results is not bad news for the Born-Infeld theory.
 But this clearly calls for a deeper inquiry. 

Work financially supported by NSF Grants 
DMS-0103808 and  DMS-0406951.

%
%
%

\thebibliography
\bibitem{}

\bibitem{efat} 
	E. S. Fradkin 
	and 
	A. A. Tseytlin, 
	Phys. Lett. \textbf{B163}, 123 (1985). 

\bibitem{mbliA} 
	M. Born 
	and 
	L. Infeld, 
	Proc. R. Soc. London \textbf{A144}, 425 (1934).

\bibitem{ggB}
	G. Gibbons, 
	Rev. Mex. Fis. \textbf{49S1}, 19 (2003). 

\bibitem{ggA}
	G. Gibbons, 
	Nucl. Phys. \textbf{B514}, 603 (1998).

\bibitem{yy}
        Y. Yang, 
        Proc. Roy. Soc.  London \textbf{A 456}, 615 (2000). 

\bibitem{dgrk}
	D. Gal'tsov and R. Kerner,
	Phys. Rev. Lett. \textbf{84}, 5955 (2000). 

\bibitem{mbA}
        M. Born, 
        Nature \textbf{132}, 282 (1933).

\bibitem{remarkA}
	In \cite{mbA} Born proposed a simpler nonlinearity than in 
	\cite{mbliA}; yet in the electrostatic limit both field theories 
	coincide.

\bibitem{swA}
        S. Weinberg, 
                \textit{$\!$The Quantum theory of fields I},
        Cambridge Univ. Press, Cambridge (1995).

\bibitem{ibb}
        I. Bia\l ynicki-Birula,         
	in: ``Quantum theory of particles and fields,'' pp. 31-48,
        in honor of Jan {\L}opusza{\'n}ski; eds.
        B. Jan\-ce\-wicz and J. Lukierski,
        World Scientific, Singapore (1983).

\bibitem{es}
        E. Schr\"odinger, 
        Proc. Roy. Irish Acad. \textbf{A 48},  91 (1942). 

\bibitem{mkA} 
	M.K.-H. Kiessling,
	J. Stat. Phys. {\bf 116}, 1057 (2004).  

\bibitem{pamd}
        P.A.M. Dirac, 
        Proc. Roy. Soc. A \textbf{257},  32 (1960). 

\bibitem{mbB}
        M. Born, 
                 \textit{Atomic physics}, $8^{\mathrm{th}}$ rev. ed.,
        Blackie \& Son Ltd., Glasgow (1969).

\bibitem{ghlm}
	G. Heller
	and
	L. Motz,
	Phys. Rev. \textbf{46}, 502 (1934). 

\bibitem{jrlfwg}
        J. Rafelski, 
	L. P. Fulcher, 
	and  
	W. Greiner,
	Phys. Rev. Lett.  \textbf{27},	958 (1971).  

\bibitem{gsjrwg}
        G. Soff, J. Rafelski, and W. Greiner,
	Phys. Rev.   \textbf{A7}, 903 (1973).

\bibitem{gmr}
        W. Greiner, B. M\"uller, and J. Rafelski, 
	 \textit{Quantum Electrodynamics of Strong Fields,}
	  Springer, New York (1985).

\bibitem{jdj}
	J.D. Jackson,
                 \textit{Classical Electrodynamics}, 
	$3^{\mathrm{rd}}$ rev. ed., Wiley \& Sons, New York (1999).

\bibitem{remarkB}
	The correct curl-free electrostatic field is given by
	$\SPvec{E} = {\cal F}_{\mathrm{BI}}(\SPvec{D})$ with
	$\SPvec{D} = \SPvec{D}_{\mathrm{C}} + \nabla\times\SPvec{G}$;
	the computation of $\nabla\times\SPvec{G}$ 
	does not seem to have been achieved yet.

\bibitem{mkB} 
	M.K.-H. Kiessling, 
	J. Stat. Phys. {\bf 116}, 1123 (2004).  

\bibitem{remarkC} 
	Our $\beta^2\, \propto a$ in  \cite{mbA}.

\bibitem{data}
{http://physics.nist.gov/PhysRefData}
\endthebibliography
\vfill\eject
\end{document}